\documentclass[preprint,aps]{revtex4}
\usepackage{amsmath}

\usepackage{graphicx}

\begin{document}
\title{Superluminal Solutions to the Klein-Gordon Equation\\
and a Causality Problem}
\author{\fbox{A.A.Borghardt}, M.A.Belogolovskii}
\email{bel@kinetic.ac.donetsk.ua}
\author{D.Ya.Karpenko}
\affiliation{Donetsk Physical and Technical Institute, National
Academy of Sciences of Ukraine,\\ Str. R.Luxemburg 72, 83114
Donetsk, Ukraine}
\date{\today}

\begin{abstract}
We present a new axially symmetric monochromatic free-space
solution to the Klein-Gordon equation propagating with a
superluminal group velocity and show that it gives rise to an
imaginary part of the causal propagator outside the light cone. We
address the question about causality of the spacelike paths and
argue that the signal with a well-defined wavefront formed by the
superluminal modes would propagate in vacuum with the light speed.
\end{abstract}
\maketitle

A long-standing question of objects or waves traveling faster than
the light speed in vacuum $c$ has received a renewed interest in
the past decade. In particular, it was shown experimentally that
light pulses can propagate with abnormally large group velocities
greater than $c$ although an interpretation of the observations
still remains a matter for debate and controversies (recent
highlights  on the problem as well as all important references on
this subject can be found in a special issue of the IEEE Journal
of Selected Topics in Quantum Electronics~\cite{review}). In this
context, it is necessary to emphasize that superliminal group
velocities at any case do not contradict the causality principle
in the special theory of relativity because, in contrast to a
widespread opinion, group velocity, like phase velocity, is not
synonymous with a signal speed (see, for instance, overviews of
the problem by Vainshtein~\cite{Vainshtein} and recently by Chiao
and Milonni~\cite{Chiao} and B\"uttiker and
Washburn~\cite{Buttiker}). A real signal should start from zero at
some instant and hence should have a well-defined wavefront. Its
arrival at a given point can occur only when the point is reached
by the front. Because it is associated with infinite frequencies,
the asymptotic behavior of a group velocity defines the front
spreading. It means that group velocity at finite frequencies is
no longer a meaningful concept and there are no restrictions on
its value which can be even infinite~\cite{Chiao1,Chiao2}. What is
important is only the high-frequency asymptote that, of course,
cannot propagate faster than $c$, an upper bound to the signal
velocity.

Below we are
concerned to cylindrically symmetrical free-space solutions of the
Klein-Gordon (KG) equation with sub- and superluminal group
velocities $v$. The former ones belong to a class of localized
diffraction-free modes that have been found for a homogeneous wave
equation~\cite{Durnin} and studied in a number of papers (see
Ref.~\onlinecite{Lunardi} and references therein). The discussion
of the latter solutions as well as their relation to the causal
Green function, up to our knowledge, has not been done yet.

The KG equation follows from the well-known relation between
energy $E$, momentum $p$, and particle rest mass $m$
\begin{equation}
\label{eq:1}
E^2=c^2(p^2+m^2c^2)
\end{equation}
through the operator transformation $E\rightarrow
i\hbar\partial/\partial t$ and $\bf{p}\rightarrow -i\hbar\nabla $.
The particle velocity is defined as a first derivative $\bf{
v}=\partial E/\partial \bf{p}$. Comparing with the basic formula
of the relativistic mechanics $E=mc^2/\sqrt{1-v^2/c^2}$, Eq.
(\ref{eq:1}) contains well-known additional solutions for
negative-energy states that were understood as antimatter (within
the classical theory it should be interpreted as a negative rest
mass). What we want to emphasize here is that this equation is
also valid for a relativistic particle with $v>c$. This statement,
of course, cannot be regarded as a proof of the existence of
tachyons, the faster-than-light objects, but only as a hint that
the corresponding wave equation may contain modes with a
superluminal $v$. Another hint comes from the physics of a free
relativistic particle. A causal propagator for the KG equation is
known to be manifestly nonvanishing outside the light
cone~\cite{Feynman} and, as it was stated in
Ref.~\onlinecite{Redmount}, the path-integral formulation of
relativistic quantum mechanics does include the contribution of
spacelike trajectories. What we do not agree with the authors of
the paper~\cite{Redmount} is that this behavior is acausal, i.e.,
backward in time in some Lorentz frames. Below we shall find a
dispersion law for superluminal solutions and show that in any
case a signal velocity formed by the modes and defined as that of
a wavefront cannot exceed the light speed in vacuum. Moreover, we
shall demonstrate that an expression for a causal propagator
outside the light cone does follow directly from the dispersion
law with superluminal velocities.

Let us look for superluminal modes within a class of axially
($\varphi $ invariant) symmetric monochromatic solutions of the KG
equation propagating in the $z$ direction
$\psi(\bf{r},t)=\phi(\rho)\exp(ik_zz-i\omega t)$, where $\rho
=\sqrt{x^2+y^2}$, $\omega $ is the angular frequency, and $k_z$ is
the axial wavenumber. For $\phi(\rho)$ we obtain the following
equation
\begin{equation}
\label{eq:2} [\Delta _{\rho}+((\omega
/c)^2-k_z^2-(mc/\hbar)^2)]\phi (\rho )=0
\end{equation}
with a two-dimensional Laplace operator $\Delta _{\rho }=\partial
^2/\partial x^2+\partial ^2/\partial y^2$ in the plane
perpendicular to the $z$-axis. The usually used solutions of Eq.
(\ref{eq:2}) are plane waves proportional to $\exp(ik_xx+ik_yy)$,
another ones are Bessel functions whose form depends on the sign
of the expression in brackets. If it is positive, i.e.,
\begin{equation}
\label{eq:3} (\omega /c)^2-k_z^2-(mc/\hbar)^2=Q^2>0,
\end{equation}
then $J_0(Q\rho )$, the Bessel function of the first kind and zero
order, is a solution of Eq. (\ref{eq:2}), and
\begin{equation}
\label{eq:4} \psi (\rho ,z,t)=J_0(Q\rho )\exp(ik_zz-i\omega t).
\end{equation}
This axially symmetric mode for a wave equation ($m=0$) was found
and realized experimentally by Durnin et al.~\cite{Durnin} and is
known now as a Bessel beam~\cite{Lunardi}. Its group velocity
$v_z=\partial \omega/\partial k_z$ is always less than $c$ but the
relating wave packet would propagate with the light speed in
vacuum (as it was shown by detailed calculations in
Ref.~\onlinecite{Lunardi} and follows from the asymptotic behavior
of $v_z$).

Another possibility, i.e.,
\begin{equation}
\label{eq:5} (\omega /c)^2-k_z^2-(mc/\hbar)^2=-q^2<0,
\end{equation}
has not been discussed yet. The corresponding solution of Eq.
(\ref{eq:2}) is a familiar modified Bessel function of the first
kind and zero order $K_0(q\rho )$ and the whole solution to the KG
equation looks as
\begin{equation}
\label{eq:6} \psi (\rho ,z,t)=K_0(q\rho )\exp(ik_zz-i\omega t).
\end{equation}
Together with the relation $k_z(\omega)=\pm\sqrt{(\omega
/c)^2+q^2-(mc/\hbar)^2}$, Eq. (\ref{eq:6}) is the main result of
the Letter. It is evident that the relating group velocity $v_z$
exceeds the value of $c$ for all available $k_z$. But for great
$\omega $ it asymptotically goes to $c$ and it means that the
signal formed by these modes will propagate with the light speed
in vacuum never destroying the causality principle.

Following the paper~\cite{BK}, we can now construct fundamental
solutions of the KG equation (Green functions) for timelike and
spacelike paths in the Minkowski 4D space. Eq. (\ref{eq:4}) with
the dispersion law (\ref{eq:3}) yields~\cite{BK} a real part of
the causal Feynman propagator $\Delta ^c$~\cite{Feynman} valid
inside the light cone
\begin{eqnarray}
\frac{c}{2\pi }\int_0^{\infty}QdQJ_0(Q\rho )\int dk_z\frac{\sin
\omega(k_z)|t|}{\omega(k_z)}\exp(ik_zz)
\nonumber\\
=\frac{1}{2\pi }\int_0^{\infty}QdQJ_0(Q\rho )J_0(\tau
\sqrt{Q^2+(mc/\hbar)^2})=\mathrm{Re}\Delta ^c(\lambda ), \nonumber
\end{eqnarray}
where $\tau=\sqrt{c^2t^2-z^2}$ is the 2D timelike interval and
$\lambda =\sqrt{c^2t^2-r^2}$ is the 4D timelike interval. From the
solution (\ref{eq:6}) with the dispersion relation (\ref{eq:5}) we
obtain (see Ref.~\onlinecite{BK} for details) the Green function
outside the light cone
\begin{eqnarray}
\frac{1}{2\pi c}\int_{mc/\hbar}^{\infty}qdqK_0(q\rho )\int
d\omega\frac{\sin k_z(\omega )|z|}{k_z(\omega )}\exp(-i\omega t)
\nonumber\\
=\frac{1}{2\pi }\int_{mc/\hbar}^{\infty}qdqK_0(q\rho
)J_0(\widetilde{\tau }\sqrt{q^2-(mc/\hbar)^2})=\frac{mc}{2\pi
\hbar\widetilde{\lambda}}K_1(mc\widetilde{\lambda }/\hbar),
\nonumber
\end{eqnarray}
where $\widetilde{\tau}=\sqrt{z^2-c^2t^2}$ is the 2D spacelike
interval, $\widetilde{\lambda }=\sqrt{r^2-c^2t^2}$ is the 4D
spacelike interval, $K_1$ is the modified Bessel function of the
first order. The latter result is just an imaginary part of the
causal Feynman propagator $\Delta ^c$~\cite{Feynman} for spacelike
paths. In the limit $\hbar \rightarrow 0$ it vanishes for all
classically forbidden trajectories.

Resuming, we worked out a novel solution to the free-space Klein-
Gordon equation propagating with a superluminal group velocity
outside the light cone. We have rejected any
conclusions~\cite{Redmount} about acausality of the spacelike
paths and argued that a signal carried by a wave packet belonging
to this class of modes propagates with the light speed in vacuum
$c$. In our opinion, there are no objections to possible
realizations of such solutions in nature. If they do exist, their
thermodynamical properties would be very exotic due to the unusual
spectrum~(\ref{eq:5}).

\end{document}